\documentclass[apjl]{emulateapj}
\usepackage{apjfonts}

\received{2008 March 5}
\accepted{2008 March 17}
\newcommand\msun{\ensuremath{M_\sun}}
\newcommand\teff{\ensuremath{T_{\rm eff}}}
\newcommand\logg{\ensuremath{\log g}}

\newcommand\kg{\mbox{$\kappa$-$\gamma$}}

\slugcomment{Accepted to the Astrophysical Journal Letters} 
\shorttitle{Pulsating Carbon-Atmosphere WDs}
\shortauthors{Montgomery et al.}

\begin{document}
\title{SDSS J142625.71$+$575218.3, a Prototype for a New Class of
  Variable White Dwarf}

\author{M.\ H.\ Montgomery\altaffilmark{1}, Kurtis A.\
  Williams\altaffilmark{1,2}, D.\ E.\ Winget\altaffilmark{1},
  Patrick Dufour\altaffilmark{3}, Steven DeGennaro\altaffilmark{1},
  James Liebert\altaffilmark{3}}

\altaffiltext{1}{Department of Astronomy, University of Texas at
  Austin, Austin, TX, USA; mikemon@astro.as.utexas.edu}
\altaffiltext{2}{NSF Astronomy \& Astrophysics Postdoctoral Fellow}
\altaffiltext{3}{Steward Observatory, University of Arizona, Tucson, AZ, USA}

\begin{abstract}
  We present the results of a search for pulsations in six of the
  recently discovered carbon-atmosphere white dwarf (``hot DQ'')
  stars. Based on our theoretical calculations, the star SDSS
  J142625.71$+$575218.3 is the only object expected to pulsate.  We
  observe this star to be variable, with significant power at 417.7~s
  and 208.8~s (first harmonic), making it a strong candidate as the
  first member of a new class of pulsating white dwarf stars, the
  DQVs. Its folded pulse shape, however, is quite different from that
  of other white dwarf variables, and shows similarities with that of
  the cataclysmic variable AM CVn, raising the possibility that this
  star may be a carbon-transferring analog of AM CVn stars. In either
  case, these observations represent the discovery of a new and
  exciting class of object.
\end{abstract}

\section{The Enigma of Hot DQ White Dwarfs}

White dwarfs (WDs) are the end stage of evolution for the vast
majority of stars in the Universe.  As the cores of former stars, WDs
provide crucial observational constraints on stellar evolutionary
models.  Traditionally, two classes of WD stars were known: those with
hydrogen-rich atmospheres (spectral type DA) and those with
helium-rich atmospheres (non-DA spectral types).  Recently
\citet{Dufour2007} announced a new class of WD with carbon-dominated
atmospheres, the ``hot DQ'' stars, several examples of which have been
found in the Sloan Digital Sky Survey \citep{Liebert2003}.

The origin of single carbon-atmosphere WDs is very uncertain.  One
proposed scenario has hot DQ stars as the progeny of stars with masses
of $9-11$\msun, massive enough to ignite carbon and form an
oxygen-neon WD with a carbon-oxygen atmosphere
\citep[e.g.][]{GarciaBerro1994,GarciaBerro1997}.  Alternatively, the
hot DQ WDs may arise from a particularly violent late thermal pulse
which burns all the hydrogen and helium \citep[e.g.,][]{Herwig1999}.

White dwarf asteroseismology is a potential avenue for studying the
parameters and interior structures of hot DQ stars, providing that
g-mode pulsations can be detected. Such pulsations have been detected
in DA, DB, and PG 1159 stars, and non-radial pulsations have led to
detailed constraints on these stars \citep[see the forthcoming review
article of][]{Winget2008}.

In this Letter, we report on the discovery of pulsations in the
carbon-atmosphere object \object{SDSS J142625.71+575218.3} (hereafter
SDSS J1426+5752), and give evidence showing that this object is either
the first known ``DQV'' WD or the first known cataclysmic variable
(CV) with a carbon-dominated spectrum.  A bulletin announcing this
discovery has been published \citep{Degennaro2008}.

\section{The Motivation for Pulsation}

There are three known classes of WD pulsators, the DAV, the DBV, and
the PG 1159 (DOV) stars. The DAVs have hydrogen-dominated spectra and
pulsate at temperatures at which hydrogen is partially ionized ($\teff
\sim 12000$~K), while the DBVs have helium-dominated spectra and
pulsate at temperatures at which helium is partially ionized ($\teff
\sim 25000$~K). Thus, it is natural to expect that the hot DQ stars,
with carbon-dominated atmospheres, will also pulsate near a \teff\
associated with a partial ionization state of carbon.

More precisely, WD pulsations are seen in instability strips, with a
high-temperature boundary (the ``blue edge'') and a low-temperature
boundary (the ``red edge''). Theoretical calculations have
traditionally been done for the blue but not the red edges of the two
instability strips; the blue edge calculation is linear, whereas the
cessation of pulsation at the red edge appears to be an intrinsically
nonlinear effect. In this Letter we focus solely on the blue edge,
since it is the most pertinent for our sample of stars.

Theory and observation have firmly established that surface partial
ionization causes pulsations in WDs. A particular
pulsation mode is driven locally when maximum pressure lags maximum
density. In models this can happen in two qualitatively different
ways: the operation of the \kg\ (``kappa-gamma'') mechanism, and
``convective driving.''  The essential feature of both driving
mechanisms is that the mode periods need to be of order the thermal
timescale, $\tau_{\rm th}$, at the base of the partial ionization
zone. In the \kg\ mechanism, driving occurs locally when the opacity
varies so that a net amount of radiative flux is still flowing into a
region at maximum compression. Early calculations of driving in WDs
focused on this mechanism, mainly due to the difficulties associated
with modeling time-dependent convection.  These calculations yielded
results in good agreement with the known pulsators, and led to the
prediction and subsequent discovery that DB stars pulsate
\citep{Winget1982}.  However, the \kg\ approach is not
self-consistent, since it ignores the response of the convection zone
to the pulsations, and the turnover times in the convection zones are
of order seconds -- short compared to the observed pulsation periods
for non-radial g-modes.

\begin{figure}[t]
  \centering{
\includegraphics[height=1.0\columnwidth,angle=-90]{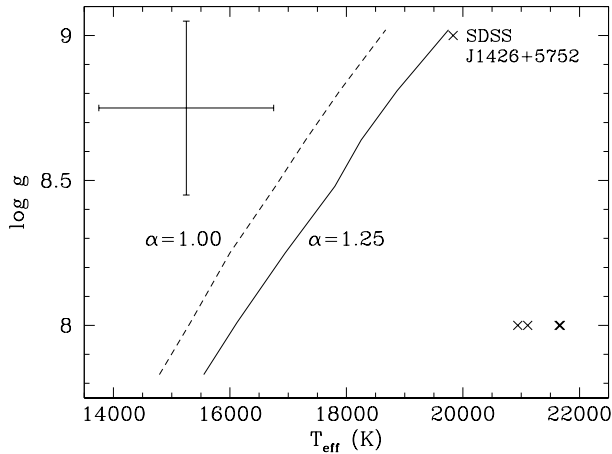}
}
\caption{A plot of the theoretical blue edge of the carbon-rich 
  instability strip in the $T_{\rm eff}$ -- $\log g$ plane, assuming
  ML2/$\alpha=1.25$ (solid line) and ML2/$\alpha=1.00$ (dashed line).
  The crosses give the current best estimates for the positions of the
  stars we observed based on the parameters in \citet{Dufour2008}; in
  the upper left-hand corner we show a representative error bar for
  these stars.}
\label{Cstrip}
\end{figure}

Improving upon this situation, \citet{Brickhill90}, and later
\citet{Goldreich99}, developed an approach which self-consistently
includes perturbations of the convective flux.  The crucial insight is
that the convection zone responds nearly instantaneously to the
pulsations, so that it is always in hydrostatic equilibrium. While
more physically sound, these calculations show that the relevant
criterion for mode driving is similar to that of the \kg\ mechanism,
namely that the periods of driven modes be of order the convective
response timescale $\tau_C$, which itself is some multiple of
$\tau_{\rm th}$.  Thus, while predictions of the location of the blue
edge of an instability strip move somewhat, generic features such as
its mass dependence and the blue edge temperature boundaries are
qualitatively unchanged.  Furthermore, similar shifts in the position
of the instability strip can be achieved through a different choice of
$\alpha$, the mixing length-to-scale height ratio, so there is not a
clear predictive difference between the two mechanisms.

Since the two mechanisms make similar predictions, we defer further
discussion of the detailed driving mechanism. We adopt the thermal
timescale at the base of the convection zone as our diagnostic, and we
assume that the blue edge corresponds to the point where $\tau_{\rm
  th} \sim 100$~s, since 100~s is at the lower end of the periods
observed in pulsating white dwarfs.

In Figure~\ref{Cstrip} we show our calculation of the location of the
blue edge of the instability strip for DQ WDs, as a function of \teff\
and $\log g$.  The depth of a WD's convection zone also
depends on the choice of the mixing length parameter $\alpha$, and we
show the results for two values of $\alpha$: $\alpha=1.25$ (solid
line) is inferred from spectroscopic fits of DBs
\citep[e.g.,][]{Beauchamp99}, and $\alpha=1.00$ (dashed line) is
inferred from nonlinear light curve fits to DBVs
\citep{Montgomery05,Montgomery07}. The range of $\log g$ chosen
corresponds to models with masses from 0.5 to 1.2 $M_{\odot}$.  The
composition, from core to atmosphere, is assumed to be pure carbon.
The stars in our sample are plotted as crosses in this figure; one
star, SDSS J1426$+$5752 lies tantalizingly close to our theoretical
blue edge.

\begin{figure}[t]
\includegraphics[angle=-90,width=\columnwidth]{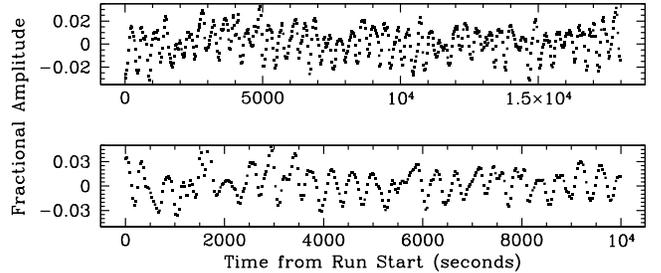}
\caption{Time-series photometry for SDSS J1426+5752, from 2008
  February 10 (top panel) and 2008 February 11 (bottom panel),
  Gaussian smoothed by 1.5 bins (45 sec). The data between $\sim
  1200$ sec and $\sim 1800$ sec in the second night were affected by a
  passing cloud.  Periodic pulsations are evident in the data from
  both nights. \label{fig.curves}}
\end{figure}

\section{Observations}

Six hot DQ stars were well-placed for observations during our
observing run in early February of 2008. Time-series photometry of
these stars were obtained on the nights of UT 2008 February 6--11 with
the Argos high-speed photometer on the McDonald Observatory 2.1m Otto
Struve Telescope \citep{Nather2004}.  Other than SDSS J1426+5752, none
of the WDs exhibited observable pulsations within the amplitude limits
of our photometry ($\approx 5$ mma). These observations will be
reported in a future paper.

Observations of SDSS J1426+5752 were obtained on the nights of UT 2008
February 10--11 We obtained continuous time series data for
uninterrupted runs of 5.0 hours (February 10) and 2.8 hr (February
11). Exposure times were 30s and taken through a Schott glass BG40
filter. The seeing ranged from 2 to 3 arcsec, and transparency
variations were minimal with the exception of a single passing cloud
during the second night's run.

The photometric data were reduced via the methods described in
\citet{Mukadam2004}. Light curves are shown in
Figure~\ref{fig.curves}. The pulsations are difficult to see in the
raw data, but readily observed when the data are smoothed with a
Gaussian filter with $\sigma= 1.5$ bins (45~sec).

\begin{deluxetable}{ccc}
\tablecolumns{3}
\tablewidth{0pt}
\tablecaption{Frequency solution for SDSS J$1426+5752$
  \label{tab.freq}}
\tablehead{\colhead{Frequency ($\mu$Hz)} & \colhead{Amplitude (mma)} &
 \colhead{Phase}}
\startdata
\multicolumn{3}{c}{\bf Combined data set} \\
\hline\\
~2394.27 $\pm$ 0.19 & 17.54 $\pm$ 0.82 & 1.00 $\pm$ 0.01 \\
~4788.82 $\pm$ 0.49 & ~6.67 $\pm$ 0.82 & 0.30 $\pm$ 0.02 \\
12053.41 $\pm$ 1.02 & ~3.20 $\pm$ 0.82 & 0.40 $\pm$ 0.04  \\
\hline\\
\multicolumn{3}{c}{\bf First Night} \\
\hline \\
~2393.25 $\pm$ ~1.60 & 16.11 $\pm$ 0.84 & 0.97 $\pm$ 0.05  \\
~4787.63 $\pm$ ~3.91 & ~6.58 $\pm$ 0.84 & 0.26 $\pm$ 0.12  \\
12056.99 $\pm$ 17.46 & ~1.47 $\pm$ 0.84 & 0.52 $\pm$ 0.52 \\ 
\hline\\
\multicolumn{3}{c}{\bf Second Night} \\
\hline\\
2393.06 $\pm$ ~4.61 & 20.07 $\pm$ 1.68 & 0.06 $\pm$ 0.25 \\
~4766.32 $\pm$ 12.25 & ~7.51 $\pm$ 1.68 & 0.48 $\pm$ 0.65 \\
12044.79 $\pm$ 14.39 & ~6.42 $\pm$ 1.68 & 0.85 $\pm$ 0.77
\enddata
\end{deluxetable}

\section{Analysis}

\begin{figure}[t]
  \centering{
\includegraphics[width=\columnwidth]{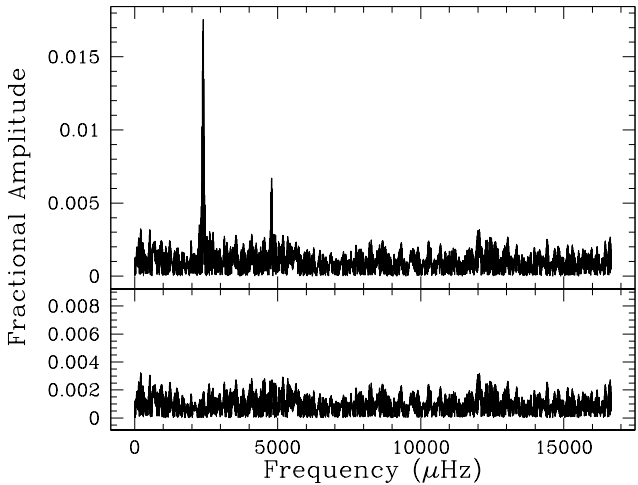}
}
\caption{Top panel: the Fourier transform of the two nights of data on
  SDSS~1426+5752.  The dominant features in the FT are the main mode
  at 417.66 s and its first harmonic. Lower panel: the Fourier
  transform after pre-whitening by the main mode and its first
  harmonic.}
  \label{FT}
\end{figure}

In the upper panel of Figure~\ref{FT} we present the discrete Fourier
transform (FT) of the unsmoothed combined data set. It shows a single
pulsational period of 417.66 sec (2394.27 $\mu$Hz) as well as the
first harmonic of this peak (208.82 sec). There is also a hint of
power around 12000 $\mu$Hz, which is near the fourth harmonic of the
main frequency though it may also be a signature of periodic drive
errors. After pre-whitening by the two main periodicities, no other
pulsational frequencies are observed with an amplitude higher than
3.5~mma (lower panel, Figure~\ref{FT}). We note that these periods are
consistent with those seen in other white dwarfs with g-mode
pulsations.

In Table~\ref{tab.freq} we summarize frequency fits of this data, for
both the combined data set and the individual nights. We have included
the main periodicity and its harmonic in these fits, as well as one at
12053.41 $\mu$Hz. This last frequency was chosen because it is the
highest in this region of the FT, but due to aliasing another nearby
peak in the FT may be the true frequency. More data are needed to
determine the frequency more accurately.

Notice that the frequencies, amplitudes, and phases are broadly
consistent within the errors; this, together with the successful
pre-whitening, show that these modes were coherent across the 2-day
baseline. We note that the amplitude of the $\sim 12050~\mu$Hz mode
changed significantly between the two nights, being much larger the
second night, and that the frequency of the $\sim 4790~\mu$HZ mode was
$\sim 2 \sigma$ lower the second night than the first.

\begin{figure}
  \centering{
\includegraphics[width=1.0\columnwidth]{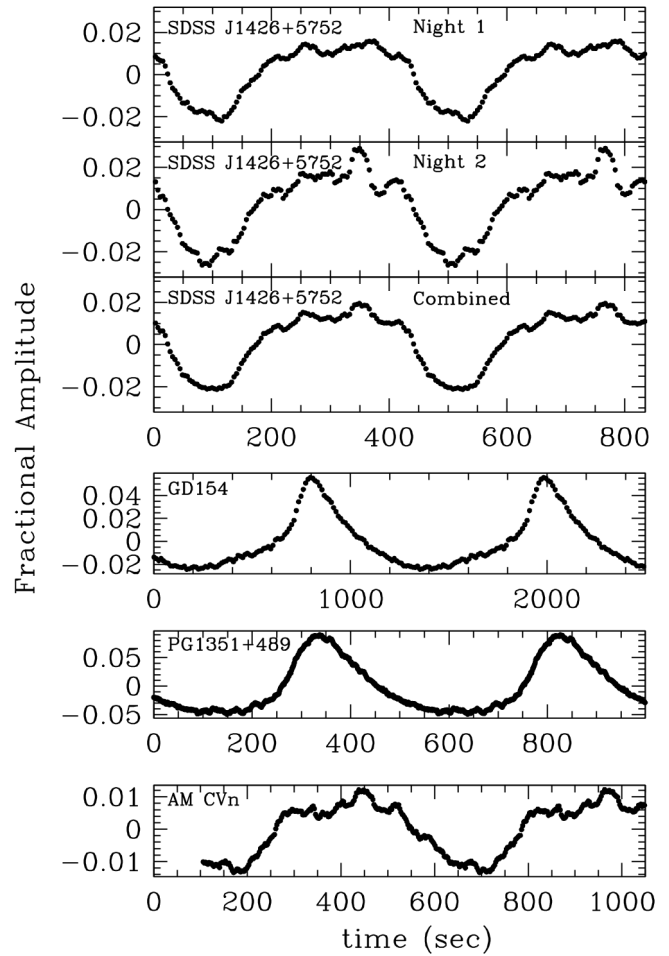}
}
 \caption{The top three panels are the light curve of SDSS J1426+5752 
   folded at a period of 417.66~s; the panels show the pulse shapes of
   the first night, second night, and both nights combined.  For
   comparison, the bottom three panels show the pulse shapes of the
   DAV GD154, the DBV PG1351+489, and the CV AM CVn (HZ 29),
   respectively.  }
\label{fold}
\end{figure}

Since the light curve of this object is dominated by one frequency, we
have computed a pulse shape by folding the data at a periodicity of
417.66 s (see Figure~\ref{fold}). This unusual pulse shape is produced
mainly by the presence of a first and a fourth harmonic.  While the
first harmonic could be due to ``normal'' nonlinear processes, such as
a varying convective response \citep[e.g., ][]{Montgomery05} or
bolometric flux corrections \citep[e.g., ][]{Montgomery08,Brassard95},
the lack of 2nd and 3rd harmonics is puzzling.

In the lower panels of Figure~\ref{fold} we show the pulse shapes of
the only known single-mode WD pulsators with stable oscillation
spectra, the DAV GD154 and the DBV PG1351+489. Even without the fourth
harmonic (e.g., see top panel in Figure~\ref{fold}, when this harmonic
had a small amplitude), the pulse shape of SDSS J1426+5752 looks
different from those of other stars. This is because the \emph{phase}
of the first harmonic is such that it makes the peaks lower and the
valleys deeper.  In a typical WD pulsator, the phase of the first
harmonic makes the peaks higher and the valleys shallower.

While unlike that of known WD pulsators, the pulse shape of SDSS
J1426+5752 \emph{is} similar to one observed in a different type of
object. In the bottom panel of Figure~\ref{fold}, we show a folded
light curve from the object AM CVn (= HZ 29), a cataclysmic variable
with a low-mass, presumably degenerate, helium star transferring mass
to a white dwarf.  The similarity to SDSS J1426+5752 is striking and
serves to muddy the interpretation of this object. However, it should
be noted that the pulse shape of AM CVn varies significantly, and at
other epochs the similarities are not nearly as pronounced
\citep[e.g., see][]{Provencal95}.

\subsection{A new class of pulsator?} 

SDSS J1426+5752 may be the first of a new class of pulsating
carbon-atmosphere WDs, the DQVs. These observations were motivated by
theoretical calculations (see Figure~\ref{Cstrip}) which indicated
that it was the only member of our sample of six hot DQ stars which
should pulsate, and it \emph{is} the only one found to vary. Its
spectrum is consistent with that of a single, massive
carbon-atmosphere white dwarf \citep{Dufour2008}. Given these facts,
it seems natural to conclude that SDSS J1426+5752 is a member of a new
class of WD pulsators.

At present, the \emph{only} difficulty with the pulsator
interpretation is the pulse shape, unique among the pulsating WD
stars. If it really is a WD pulsator, then the modes with frequencies
near the first and fourth harmonic must be independent pulsation
modes, and therefore free to have arbitrary phases. The reason these
modes are excited would then be due to a parametric resonance with the
main mode \citep[e.g., see][]{Goupil94}. The likelihood of resonances
is increased by the high mass of this star (as preferred by the
spectral models), since the density of modes increases with the mass of
the star.  Further, there is some evidence in the light curve of
beating with a period of $\sim 4000$~s, though the noise level
prevents meaningful conclusions about the reality of a mode with this
frequency separation.  We await further observations.

\subsection{An interacting double-degenerate?} 

A few characteristics of SDSS J1426+5752 are similar to \object{AM
CVn} stars.  The light curves of several AM CVn systems show periodic
variations consistent with a fundamental frequency and multiple
harmonics.  The folded light curve of AM CVn itself is qualitatively
similar to that for SDSS J1426+5752 \citep[see Figure~\ref{fold},
and][]{Provencal95}.  This opens the possibility that SDSS J1426+5752
may be an AM CVn-like system, but with a carbon-dominated donor star.  Such
a system could make a compelling Type Ia supernova progenitor,
depending on the total system mass.

In this scenario, SDSS J1426+5752 is a binary system consisting of two
carbon-oxygen WDs, driven close together by gravitational radiation
until mass transfer initiated, so the observed spectrum would be from
an optically-thick carbon-oxygen accretion disk around the more
massive WD star.  The spectra of high-state AM CVn systems show broad
absorption features from an optically-thick disk
\citep[e.g.,][]{Warner1995}, which would mimic high \logg\ such as
that claimed for SDSS J1426+5752.

However, models of mass transfer between two carbon-oxygen WDs suggest
that the secondary star is disrupted in only a few dynamical times and
would not evolve into an AM CVn-like system
\citep[e.g.,][]{Benz1990,Rasio1995}.  And although a thick accretion
disk may possibly remain around the primary WD for $\sim 10^6$ yr
after the disruption of the secondary \citep{Piersanti2003}, the
observed harmonics (explained in AM CVn systems as disk ellipticities,
precession, and bright spots) would be hard to explain without a
companion.  Further, oxygen, which should be present in this scenario,
has not yet been detected in the hot DQs, although the oxygen
abundance limits are
weak.

\section{Conclusions}

We have conducted a search for pulsations in six of the recently
discovered DQ stars \citep{Dufour2007}. Based on our theoretical
calculations, the star SDSS J1426+5752 was the only object predicted
to pulsate, and it \emph{is} the only target observed to be variable.
This is a strong argument that SDSS J1426+5752 is the first member of
a new class of pulsating white dwarf stars, the DQVs.

Another possibility, however, is that SDSS J1426+5752 is a carbon
analog of an AM CVn system.  Currently, the only evidence in favor of
this model is the similarity of its pulse shape with AM CVn;
we consider this possibility less likely.

How can we distinguish between these models?  Signatures of disk
activity, such as flickering on very short timescales, and
radial-velocity variations in the observed spectra were necessary to
determine the nature of AM CVn. Additional observations, both time
series photometry and higher-quality spectra, are therefore necessary
to determine if this system is an interacting binary.

Both of these possibilities imply that SDSS J1426+5752 is a prototype
for a new class of white dwarf: either a pulsating carbon-atmosphere
white dwarf or a carbon-dominated AM CVn-like system. The former would
signal the discovery of a new class of pulsating WD, the first in over
25 years, while the latter could be a compelling candidate for a Type
Ia supernova progenitor.  We will continue the search for other
objects of this exciting and enigmatic class.

\acknowledgements M.H.M., K.A.W., and S.D. are grateful for the
financial support of the National Science Foundation, under awards
AST-0507639, AST-0602288, and AST-0607480, respectively; J.L.\ and
P.D.\ are grateful for the support of NSF award AST-0307321. M.H.M.
gratefully acknowledges the support of the Delaware Asteroseismic
Research Center. D.E.W. is a fellow of CNPq, Brasil, and gratefully
acknowledges their support.  P.D. acknowledges the financial support
of NSERC. The authors are grateful to R.E.  Nather and A. Mukadam for
the development and testing of the Argos instrument, without which
these observations would not have been possible.
 
{\it Facilities:} \facility{Struve (Argos)}

\end{document}